\documentstyle[preprint,tighten,aps,epsf]{revtex} 

\begin{document}

\title{Field theory of self-avoiding walks in random media}
       
\author{A V Izyumov and K V Samokhin$^\dagger$}

\address{Cavendish Laboratory, Madingley Road, Cambridge, CB3 0HE, UK}
  
\date{\today}

\maketitle

\begin{abstract}
Based on the analogy with the quantum mechanics of a particle propagating in a 
{\em complex} potential, we develop a field-theoretical description of the 
statistical properties  of a self-avoiding polymer chain in a random 
environment. We show that the account of 
the non-Hermiticity of the quantum Hamiltonian results in a qualitatively 
different structure of the effective action, compared to previous studies.
Applying the renormalisation group analysis, we find a transition between
the weak-disorder regime, where the quenched randomness is irrelevant, and
the strong-disorder regime, where the polymer chain collapses. However, 
the fact that the renormalised interaction constants and the chiral symmetry 
breaking regularisation parameter flow towards strong coupling raises 
questions about the applicability of the perturbative analysis. 
\end{abstract}

\newpage

\section{Introduction}
\label{introduction}

The problem of a polymer in a random environment is among the most interesting
in statistical physics. It has been known for a long time that the mean 
square end-to-end distance of a pure self-avoiding walk (SAW) of length $L$ 
obeys the scaling law $\langle r^2\rangle\sim L^{2\nu}$, where $\nu\approx 
0.59$ in three dimensions \cite{DeGen72,Emery75,ZJ96,KBB81}
(for a Gaussian random walk, one has the classical exact result $\nu=0.5$ 
in all dimensions).
The question of how this scaling behaviour is affected by external impurities 
has attracted considerable research effort for more than a decade \cite{Krem81,Harr83,Kim83,BM87,EM88,CB88,EC88,Thir88,MK90,Obukh90,LDM91,Step92,HV94}. 
In his pioneering work, Harris \cite{Harr83} argued that, treated 
perturbatively, quenched disorder is irrelevant, and, therefore, no 
modification of the critical exponent $\nu$ should be expected (see also
Ref. \cite{Kim83}). 
This conclusion found support in the Monte-Carlo simulations on weakly diluted 
lattices \cite{Krem81}. The opposite case of strong disorder has also been 
studied, both numerically \cite{BM87} and analytically \cite{EM88,CB88}.
Edwards and Muthukumar \cite{EM88} and Cates and Ball \cite{CB88} 
predicted that a Gaussian chain placed in the field of impurities 
would collapse to a localised state, 
in which $\langle r^2\rangle\to{\rm const}$ at $L\to\infty$. 
The localisation breaks down if one introduces a weak repulsive interaction 
between the monomers in the chain. In this case, it was found 
\cite{EC88} that the polymer behaves as a free random walk with 
$\nu=0.5$. Later, a crossover between the regimes of weak 
($\nu\approx 0.59$) and strong ($\nu=0.5$) disorder was 
predicted to occur at some critical concentration of impurities 
\cite{Thir88}. However, these results seem to contradict the conclusions 
of a number of other authors \cite{Obukh90,LDM91}, who argued that,
at any concentration of impurities, the scaling of a long chain with 
excluded volume interactions is controlled by the critical exponent of a 
directed random walk, $\nu = 2/3$. Overall, a comprehensive theory that would 
describe the effects of disorder on self-avoiding polymers is still missing.
 
Previous studies of the interplay of  disorder and excluded volume 
interactions have been largely based on qualitative arguments,
although a number of mathematical techniques (including the variational 
methods \cite{EC88,Thir88}, replica field theories \cite{Kim83,MK90,Obukh90} 
and the real-space renormalisation group \cite{Step92}) have also been used. 
A more general framework would therefore be beneficial to a better 
understanding of these effects. We believe that such a framework could be 
designed by employing a field-theoretical approach, which draws on 
the connection between the statistical mechanics of a polymer and the 
quantum mechanics of a particle in a {\em complex} random potential. 
Despite its apparent simplicity, the use of this method has been hampered 
by the fact that, as the Hamiltonian of the particle in a complex potential 
is non-Hermitian, it is impossible to represent the Green function in the
form of a convergent functional integral (see below). A similar 
problem has been encountered, and successfully resolved, in the spectral 
theory of non-Hermitian operators. The latter has attracted great 
interest in recent years. A variety of applications have been identified 
including the study of anomalous diffusion in random media \cite{Isich92}, 
scattering in open quantum systems \cite{FS97}, neural networks \cite{SCS88}, 
and the statistical mechanics of flux lines in superconductors \cite{HN96}. 
The problem of a self-avoiding walk without impurities, which can be 
mapped onto the quantum mechanics of a particle propagating in a random 
{\em imaginary} potential, has also been analysed in this context \cite{IS99}.
These studies have led to the development of a new technique,
based on the representation of the spectral properties of non-Hermitian 
operators through an auxiliary Hermitian operator of twice the dimension 
\cite{FZ97}, which serves as a good starting point for a field-theoretical 
approach. 

The main purpose of the present paper is to derive a consistent 
field-theoretical formulation of the problem of a self-avoiding polymer 
chain in a random white-noise potential, using the methods of
non-Hermitian quantum mechanics, which is done in Section \ref{derivation}.
The large-scale behaviour of this model is studied perturbatively in Section \ref{rg} 
by means of the momentum-shell renormalisation group (RG)
in $D=4-\epsilon$ dimensions. Section \ref{discussion} concludes with
a discussion of the results obtained and of possible limitations on the
validity of the perturbative approach. 

\section{Derivation of the field theory}
\label{derivation}

Let us consider a continuum self-avoiding chain of length $L$, with one end 
fixed at the origin. Then the probability to find the other end at a point 
${\bf r}$ can be expressed as a path integral (Edwards model 
\cite{Edw65,Klein95}):
\begin{eqnarray}
\label{edwards}
 P({\bf r},L)=\int_{{\bf x}(0)={\bf 0}}^{{\bf x}(L)={\bf r}} 
   {\cal D}{\bf x}(s)\;\exp\biggl\{&&-\frac{1}{2a}\int_0^L ds\;
    \left(\frac{d{\bf x}(s)}{ds}\right)^2-\int_0^Lds\;V_1({\bf x}(s)) 
  \nonumber\\
  &&-\frac{\gamma_2}{2}\int_0^Lds_1\int_0^Lds_2\;\delta({\bf x}(s_1)-
    {\bf x}(s_2))\biggr\}.
\end{eqnarray}
The first term in the exponent corresponds to the entropic contribution (the
length $a$, called the Kuhn length, is a microscopic 
parameter with the physical meaning of the monomer size, which provides a 
natural ultraviolet cutoff scale). The second term is the potential energy of 
the chain in an external potential, which is assumed to be a Gaussian 
distributed random function with the correlator 
$\overline{V_1({\bf r})V_1({\bf r}')}=\gamma_1\delta({\bf r}-{\bf r}')$. 
The last term takes into account the excluded volume effects 
($\gamma_2>0$), the limit $\gamma_2\to\infty$ describing the situation where 
the intersections of different fragments of the chain are penalized by an 
infinite energy barrier and thus completely forbidden. This interaction term 
can be decoupled by introducing an auxiliary white-noise potential. 
Using the identity
$$
 \exp\left\{-\frac{\gamma_2}{2}\int_0^Lds_1\int_0^Lds_2\;\delta({\bf x}
 (s_1)-{\bf x}(s_2))\right\}=\left\langle\exp\left\{-i\int_0^Lds\;
 V_2({\bf x}(s))\right\}\right\rangle,
$$
where the angular brackets denote averaging over a Gaussian distributed 
random field $V_2({\bf r})$ with correlator $\langle V_2({\bf r})
V_2({\bf r}')\rangle=\gamma_2\delta({\bf r}-{\bf r}')$, we can represent 
Eq. (\ref{edwards}) as an averaged Feynman propagator of a fictitious 
quantum particle moving in a {\em complex} random potential: 
$P({\bf r},L)=\langle U({\bf r},L)\rangle$, where 
$\hat U (L)=\exp(-L\hat H)$ with the Hamiltonian
\begin{equation}
\label{hamilt}
 \hat H=-\nabla^2+V_1({\bf r})+iV_2({\bf r})
\end{equation}
($V_{1,2}({\bf r})$ are independent Gaussian random fields). Note that $\hat H$
is non-Hermitian and may therefore have complex eigenvalues. 

It is convenient to use the ``energy representation'', which is 
achieved by introducing the Green operator as a function of $z=x+iy$,
\begin{equation}
\label{g(z)}
 \hat g(z)\equiv{1\over z-\hat H}=\sum_k |R_k\rangle {1\over z-z_k} 
 \langle L_k|,
\end{equation}
where $|R_k\rangle$ and $\langle L_k|$ are the right and left eigenfunctions 
of $\hat H$, and $z_k$ denote the complex eigenvalues. Using the identity
$\partial z^{-1}/\partial z^*=\pi\delta^2(z)\equiv\pi\delta(x)\delta(y)$, 
one can relate $\hat U (L)$ to $\hat g (z)$, and express the end-to-end 
probability distribution of a self-avoiding chain in a given distribution
of impurities in the form
\begin{equation}
 P({\bf r},L|V_1)={1\over \pi}\int d^2z\;\exp(-zL)
 \frac{\partial}{\partial z^*}\langle g({\bf r},z)\rangle,
\end{equation}
where the integration runs over the entire complex plane. It should be noted
that the standard field-theoretical methods, based on the representation 
of an inverse matrix in the form of a Gaussian functional integral, 
cannot be directly used for the calculation of $\hat g(z)$, the reason being 
that, as $\hat H$ may have eigenvalues anywhere in the complex plane, it is 
impossible to guarantee the convergence of the functional integral. As pointed
out in Ref. \cite{SCSS88}, a naive attempt to calculate the functional 
integral by the analytical continuation from the region where it converges 
to the whole complex plane fails. The problems are revealed by representing 
the density of complex eigenvalues through the identity
\begin{equation}
 \rho(z)\equiv\sum_k\delta^2(z-z_k)={1\over \pi}{\partial\over \partial z^*}
 \,{\rm Sp}\,\hat g(z),
\end{equation}
wherein the Green function is shown to be non-analytic everywhere in 
which the density of states is non-vanishing. 

To circumvent these difficulties, a representation has been introduced 
\cite{FZ97,Efetov97,CW97}, in which the complex Green function 
$\hat g(z)$ is expressed through an auxiliary {\em Hermitian} operator, which
in our case has the form
\begin{equation}
\label{G_matrix}
\hat G^{-1}(z)\equiv\pmatrix{ 0 & z-\hat H\cr z^*-\hat H^\dagger 
 & 0 \cr}=\left(x+\nabla^2-V_1({\bf r})\right)\sigma_1-(y-V_2({\bf r}))
\sigma_2,
\end{equation}
where $\sigma_i$ are the Pauli matrices. A relationship between $\hat g$ 
and $\hat G$ is straightforward:
\begin{eqnarray}
\label{g}
\hat g(z)=\hat G^{21}(z).
\end{eqnarray}
Using the replica trick, the matrix Green function $\hat G(z)$ can be 
written as a functional integral over $2n$-component complex Bose fields 
(in the limit $n\to 0$):
\begin{equation}
\label{G_R}
 G^{ij}({\bf r},z)=-i\lim\limits_{\eta\to +0}\int{\cal D}^2
 \mbox{\boldmath $\varphi$}_a\;\exp\left\{i\int d^Dr\;
 \mbox{\boldmath $\varphi$}_a^\dagger(\hat G^{-1}(z)+i\eta\sigma_0)
 \mbox{\boldmath $\varphi$}_a\right\}\varphi^i_1({\bf r})
 \varphi^{j,*}_1({\bf 0})
\end{equation}
($\sigma_0$ is the unit matrix, summation over repeated replica indices is 
assumed), where
$$
 \mbox{\boldmath $\varphi$}_a({\bf r})={\varphi_a^1({\bf r})\choose
 \varphi_a^2({\bf r})}, \quad
 {\cal D}^2\mbox{\boldmath $\varphi$}_a= 
 \prod\limits_{a=1}^n \prod_{i=1,2}
 \frac{{\cal D}({\rm Re}\,\varphi_a^i){\cal D}({\rm Im}\,\varphi_a^i)}{\pi}.
$$
Due to the Hermiticity of $\hat G^{-1}(z)$ and the presence of the term with 
$\eta$ (``regulator''), the functional integral is well defined and 
convergent. Note that, although the matrix Green function (\ref{G_matrix}) 
possesses a chiral symmetry, $\sigma_3\hat G^{-1}(z)\sigma_3=-\hat G^{-1}(z)$,
this symmetry is broken in Eq. (\ref{G_R}) if $\eta\neq 0$.  

Eqs. (\ref{g}) and (\ref{G_R}) allow one to average $\hat g(z)$ over $V_2$ 
and calculate the polymer partition function in a given configuration of 
the external disorder. In order to obtain physically observable quantities, 
one has to average the free energy $F({\bf r},L)=\ln P({\bf r},L)$ over the 
quenched random field $V_1$, which can be done using the replica trick 
once again:
\begin{eqnarray}
\label{F_av}
 \overline{F({\bf r},L)}&=&\lim\limits_{m\to 0}\frac{\overline{ 
 P^m({\bf r},L)}-1}{m}\nonumber\\
 &=&\lim\limits_{m\to 0}\frac{1}{m}\left\{\int\prod\limits_{\alpha=1}^m
 \frac{d^2z_\alpha}{\pi}\;\exp\left(-L\sum\limits_{\alpha=1}^m
 z_\alpha\right)
 \prod\limits_{\alpha=1}^m\frac{\partial}{\partial z^*_\alpha}\,\overline{
 \prod\limits_{\alpha=1}^m\langle g({\bf r},z_\alpha)\rangle}-1\right\}.
\end{eqnarray}
It is thus necessary to introduce a second set of replica indices and 
to integrate over $2nm$-component fields $\varphi^i_{a,\alpha}({\bf r})$ 
($i=1,2$; $a=1,...,n$; $\alpha=1,...,m$). Averaging over $V_1$, we obtain from
Eqs. (\ref{g}) and (\ref{G_R}):
\begin{equation}
 \overline{\prod\limits_\alpha\langle g({\bf r},z_\alpha)\rangle}=
 \int{\cal D}^2\mbox{\boldmath $\varphi$}_{a,\alpha}\;
 e^{iS[\mbox{\boldmath $\varphi$}_{a,\alpha}^\dagger,
 \mbox{\boldmath $\varphi$}_{a,\alpha}]}\,\prod\limits_{\alpha}
 \varphi^2_{1,\alpha}({\bf r})\varphi^{1,*}_{1,\alpha}({\bf 0})\quad
 (n,m\to 0),
\end{equation}
where the effective action has the following form:
\begin{eqnarray}
\label{action}
 iS[\mbox{\boldmath $\varphi$}]=\int d^Dr\;\biggl\{&&
 i(\mbox{\boldmath$\varphi$}_{a,\alpha}^\dagger\sigma_1\nabla^2\mbox{\boldmath
 $\varphi$}_{a,\alpha})+ix_\alpha(\mbox{\boldmath$\varphi$}_{a,\alpha}^\dagger
 \sigma_1\mbox{\boldmath$\varphi$}_{a,\alpha})-iy_\alpha(\mbox{\boldmath
 $\varphi$}_{a,\alpha}^\dagger\sigma_2\mbox{\boldmath$\varphi$}_{a,\alpha})
 -\eta(\mbox{\boldmath$\varphi$}_{a,\alpha}^\dagger\sigma_0
 \mbox{\boldmath$\varphi$}_{a,\alpha})\nonumber\\
 &&-\frac{\gamma_1}{2}(\mbox{\boldmath$\varphi$}_{a,\alpha}^\dagger\sigma_1
  \mbox{\boldmath$\varphi$}_{a,\alpha})(\mbox{\boldmath
  $\varphi$}_{b,\beta}^\dagger\sigma_1\mbox{\boldmath$\varphi$}_{b,\beta})
 -\frac{\gamma_2}{2}(\mbox{\boldmath$\varphi$}_{a,\alpha}^\dagger\sigma_2
 \mbox{\boldmath$\varphi$}_{a,\alpha})(\mbox{\boldmath
 $\varphi$}_{b,\alpha}^\dagger\sigma_2\mbox{\boldmath$\varphi$}_{b,\alpha})
 \nonumber\\
 &&-\frac{\gamma_1'}{2}(\mbox{\boldmath$\varphi$}_{a,\alpha}^\dagger\sigma_1
 \mbox{\boldmath$\varphi$}_{a,\alpha})(\mbox{\boldmath
 $\varphi$}_{b,\alpha}^\dagger\sigma_1\mbox{\boldmath$\varphi$}_{b,\alpha})
 \biggr\}.
\end{eqnarray}
This action is different from the replica $n$-vector model, used previously 
in the RG analysis of the problem of a SAW in a random medium 
\cite{Kim83,MK90,Obukh90}, because of the double dimensionality of the fields 
involved and a larger number of the coupling constants (three instead of 
two). The additional term proportional to $\gamma_1'$ cannot be obtained in 
a formal derivation of Eq. (\ref{action}), but should be added to the 
long-wavelength effective action for consistency, as we shall see below. 
Note also that we have kept the term with $\eta$ in Eq. (\ref{action}) (the 
importance of this term will become clear shortly). 
The asymmetry between the ways the Latin and Greek replica indices appear 
in Eq. (\ref{action}) reflects the differences in the nature of
the random potentials $V_{1,2}$: $V_1$ describes the external quenched disorder,
while $V_2$ is a fictitious annealed random field.

\section{Renormalisation group analysis}
\label{rg}

The long-distance properties of the field theory (\ref{action}) can be 
investigated using the momentum shell RG approach and the $\epsilon$-expansion
near the upper critical dimension $D_c=4$. Using the standard procedure,
consisting of the separation of ``slow'' and ``fast'' degrees of freedom
followed by a rescaling of lengths and fields \cite{CL95}, we obtain
the flow equations in one-loop order:
\begin{eqnarray}
\label{RG_result}
 && \frac{d\ln x_\alpha}{d\xi}=2+\frac{1}{8\pi^2}(\gamma_1-\gamma_2+
    \gamma_1'),\nonumber\\
 && \frac{d\ln y_\alpha}{d\xi}=2+\frac{1}{8\pi^2}(\gamma_1-\gamma_2+
    \gamma_1'),\nonumber\\
 && \frac{d\ln\eta}{d\xi}=2+\frac{1}{8\pi^2}(\gamma_1+\gamma_2+
    \gamma_1'),\nonumber\\
 && \frac{d\gamma_1}{d\xi}=\epsilon\gamma_1+\frac{1}{4\pi^2}(2\gamma_1^2-
 \gamma_1\gamma_2+\gamma_1\gamma_1'),\\
 && \frac{d\gamma_2}{d\xi}=\epsilon\gamma_2+\frac{1}{4\pi^2}(3\gamma_1\gamma_2
 -\gamma_2^2+3\gamma_2\gamma_1')\nonumber,\\
 && \frac{d\gamma_1'}{d\xi}=\epsilon\gamma_1'+\frac{1}{4\pi^2}
 (3\gamma_1\gamma_1'+\gamma_2^2-\gamma_2\gamma_1'+2\gamma_1^{\prime 2})
 \nonumber,
\end{eqnarray}
where $\xi=\ln (L/a)$ is the logarithmic RG parameter. The real and
imaginary components of the complex ``energy'' $z=x+iy$ are renormalised in the
same way. It also follows from Eqs. (\ref{RG_result}) that, even though 
initially $\gamma'_1=0$, it acquires a non-zero value under renormalisation.
Introducing a different set of independent variables,
$$
 u=\gamma_1-\gamma_2+\gamma_1',\qquad v=\gamma_2,\qquad w=\gamma_1',
$$
we rewrite Eqs. (\ref{RG_result}) as
\begin{eqnarray}
 && \label{xy} \frac{d\ln x}{d\xi}=2+\frac{1}{8\pi^2}u,\qquad 
    \frac{d\ln y}{d\xi}=2+\frac{1}{8\pi^2}u,\\ 
 && \label{eta} \frac{d\ln\eta}{d\xi}=2+\frac{1}{8\pi^2}(u+2v),\\
 && \label{u} \frac{du}{d\xi}=\epsilon u+\frac{1}{2\pi^2}u^2,\\
 && \label{v} \frac{dv}{d\xi}=\epsilon v+\frac{1}{4\pi^2}(3uv+2v^2),\\
 && \label{w} \frac{dw}{d\xi}=\epsilon w+\frac{1}{4\pi^2}(3uw+v^2+2vw-w^2). 
\end{eqnarray}
Note that the flow equations for $x,y$ and $u$ are decoupled from 
those for $v$ and $w$. 

Let us neglect the renormalisation of $\eta$, i.e. put $\eta_0=0$ (the 
consequences of allowing for $\eta_0\neq 0$ will be discussed below). 
As seen from Eq. (\ref{F_av}), the main contribution to the average 
free energy at $L\to\infty$ comes from $x,y\sim 1/L$, which provide the 
initial conditions for the equations (\ref{xy}). The scale $R_c$, at which the 
renormalised values of $x$ and $y$ become of the order of unity, represents 
the correlation length of our field theory and should be identified with 
the average size of a polymer of length $L$. Integrating Eqs. (\ref{xy}) 
with respect to $\xi$ from $\xi=0$ to $\xi_c=\ln(R_c/a)$, we obtain 
the equation which relates the polymer size to $L$:
\begin{equation}
\label{R_c}
 \ln\frac{L}{a}=\int_0^{\xi_c}d\xi\;\left(2+\frac{u(\xi)}{8\pi^2}\right).
\end{equation}
The behaviour of $u(\xi)$ essentially depends on the bare values of the 
coupling constants $\gamma_{1,2}$ (the bare value of $\gamma'_1$ is zero). 
One should distinguish between the two possibilities: In the weak 
disorder regime ($\gamma_1<\gamma_2$, i.e. $u_0<0$), $u(\xi)$ flows towards 
the stable fixed point $u^*=-2\pi^2\epsilon$. After substitution in Eq. 
(\ref{R_c}), we obtain $R_c(L)\sim L^\nu$ with $\nu\approx 1/2+\epsilon/16$. 
This exponent coincides with the result for a pure SAW 
\cite{DeGen72} which means that weak disorder is irrelevant for the 
large-scale properties of polymers. This result is in agreement with Refs.
\cite{Harr83,Kim83}.

In the strong disorder regime ($\gamma_1>\gamma_2$, i.e. $u_0>0$), the 
solution of Eq. (\ref{u})
\begin{equation}
 u(\xi)=u_0\frac{e^{\epsilon\xi}}{1-\frac{u_0}{2\pi^2\epsilon}
 (e^{\epsilon\xi}-1)}
\end{equation}
has a pole at $\xi=\xi^*$, where 
\begin{equation}
 \xi^*=\frac{1}{\epsilon}\ln\left(1+\frac{2\pi^2\epsilon}{u_0}\right).
\end{equation}
Substituting $u(\xi)\sim (\xi^*-\xi)^{-1}$ in Eq. (\ref{R_c}), we find that,
at $L\to\infty$, $R_c(L)\to ae^{\xi^*}$. Thus, the polymer size tends
to some constant independent of its length. Although this conclusion coincides
with the results of Refs. \cite{EM88,Step92,HV94},
the applicability of our one-loop RG equations in this strong-coupling regime 
is limited. 

\section{Conclusions and discussion}
\label{discussion}

Our results show that a proper account of the non-Hermiticity
leads to significant changes in the structure of the effective field theory
for self-avoiding walks in random media. The application of the one-loop 
renormalisation group analysis to the action with the doubled number of the 
degrees of freedom allowed us to predict the existence of a phase 
transition between the weak disorder regime (in which the scaling behaviour 
of the polymer is not affected by the external randomness), and the strong 
disorder regime (in which the polymer collapses into a localised state). 
This conclusion seems to be in agreement with one of the scenarios 
considered previously, notably in Refs. \cite{Kim83,Step92}. 

However, our RG calculations also indicate that there might be some hidden 
flaws within the perturbative approach, one of which is revealed by taking 
into consideration the infinitesimal regulator $\eta$, which 
explicitly breaks the chiral symmetry of Eq. (\ref{G_matrix}). It follows 
from Eq. (\ref{v}) that, in the weak disorder case, $v(\xi)$ has a pole at 
$\xi=\xi_v^*$, where 
\begin{equation}
 \xi_v^*=\frac{1}{\epsilon}\ln\left(1+2\pi^2\epsilon\frac{4v_0-|u_0|}{(2v_0-
  |u_0|)^2}\right).
\end{equation}   
After substitution in Eq. (\ref{eta}) and integration over $\xi$, we find 
that $\eta(\xi)$ grows under renormalisation, $\eta(\xi)\simeq\eta_0
(1-\xi/\xi_v^*)^{-\epsilon/2}$, and diverges at some scale $R_{c,\eta}\sim 
ae^{\xi_v^*}$. Although $\eta$ does not appear explicitly in the flow 
equations for $x,y$ and $u$, its singular RG behaviour may be an evidence
of the spontaneous breaking of the chiral symmetry \cite{Lerner99}. 
In the field-theoretical language, it would manifest itself as the appearance 
of the anomalous pairings such as $\langle\varphi^{1,*}\varphi^1\rangle$ and 
$\langle\varphi^{2,*}\varphi^2\rangle$, which could change the long-distance 
behaviour of our system at scales larger than $R_{c,\eta}$. It might be the
case that a seemingly trivial technical trick of doubling the degrees of 
freedom actually reveals deep physics related
to the presence of the additional continuous symmetries in the system, whose
effects would otherwise be completely lost. 

Furthermore, despite the presence of a stable fixed point in the RG equation 
(\ref{u}) for $u$ in the weak disorder regime, the coupling constants 
$\gamma_1$, $\gamma_2$ and $\gamma'_1$ are all renormalised to infinity at 
large scales (see Eqs. (\ref{v}) and (\ref{w})). The existence of the 
underlying RG flows towards a strong coupling regime can cast doubts in the 
validity of the scaling description itself (the authors of Refs. 
\cite{Obukh90,LDM91} suggested that the behaviour of the system could instead 
be determined by a strong-coupling fixed point, which is inaccessible by means
of the perturbation theory). It is interesting that an analogous situation 
takes place even in the absence of an external disorder 
(i.e. at $\gamma_1=0$), in which case $\gamma_2$ and $\gamma'_1$ 
both go to infinity, but $\gamma_2-\gamma'_1\to 2\pi^2\epsilon$. 
One of the possible explanations is that an important role
could be played by the so-called ``tail states'' in the spectrum of 
the Hamiltonian (\ref{hamilt}), which would significantly change the 
observable large-scale behaviour of the system. The properties and 
physical manifestations of the ``tail states'' of Hermitian operators have 
been extensively studied since the pioneering works of I.~M.~Lifshitz 
\cite{Lif64}. A generalisation of this theory to the case of non-Hermitian 
random operators presents an interesting theoretical challenge with numerous 
potential applications.  

Finally, we would like to emphasize that, since at every step of the 
calculations we deal with convergent functional integrals, our 
field-theoretical model of SAWs in random media is not only 
capable of reproducing some of the previous results on the perturbative level,
but also seems to suit well for the study of non-perturbative effects, 
such as the chiral symmetry breaking and/or the contribution of the spectral 
``tails''. We leave the investigation of such effects for future work.

\section*{Acknowledgements}

The authors would like to thank B.~Simons for numerous stimulating 
discussions, I.~Lerner for insightful comments and communicating to us his
unpublished results, and A.~Moroz for interest to this work. 
K.~V.~S. is also pleased to thank A.~Comtet and S.~Nechaev for interesting 
discussions and suggestions. 
The financial support by the Engineering and Physical Sciences Research 
Council, UK (K.~V.~S.) and Trinity College, Cambridge, UK (A.~V.~I.) 
is gratefully acknowledged.


\begin{references}

\bibitem[\dagger]{byline} Permanent address: L. D. Landau Institute 
  for Theoretical Physics, Kosygina Str. 2, 117940 Moscow, Russia.\\

\bibitem{DeGen72} 
De Gennes P G 1972
{\em Phys. Lett.} {\bf 38}A 339

\bibitem{Emery75}
Emery V J 1975
{\em Phys. Rev.} B {\bf 11} 239

\bibitem{ZJ96}
Zinn-Justin J 1996 
{\em Quantum Field Theory and Critical Phenomena} (Oxford: Clarendon Press)

\bibitem{KBB81}
Kremer K, Baumg\"artner A and Binder K 1981
{\em Z. Phys.} B {\bf 40} 331

\bibitem{Krem81}
Kremer K 1981
{\em Z. Phys.} B {\bf 45} 149

\bibitem{Harr83}
Harris A B 1983
{\em Z. Phys.} B {\bf 49} 347

\bibitem{Kim83}
Kim Y 1983
{\em J. Phys. C: Solid State Phys.} {\bf 16} 1345

\bibitem{BM87}
Baumg\"artner A and Muthukumar M 1987 
{\em J. Chem. Phys.} {\bf 87} 3082

\bibitem{EM88}
Edwards S F and Muthukumar M J 1988
{\em J. Chem. Phys.} {\bf 89} 2435

\bibitem{CB88}
Cates M E and Ball R C 1988 
{\em J. Physique} {\bf 49} 2009

\bibitem{EC88}
Edwards S F and Chen Y 1988
{\em J. Phys. A: Math. Gen.} {\bf 21} 2963

\bibitem{Thir88}
Thirumalai D 1988
{\em Phys. Rev.} A {\bf 37} 269

\bibitem{MK90} 
Machta J and Kirkpatrick T R 1990 
{\em Phys. Rev.} A {\bf 41} 5345

\bibitem{Obukh90}
Obukhov S P 1990 
{\em Phys. Rev.} A {\bf 42} 2015

\bibitem{LDM91}
Le Doussal P and Machta J 1991 
{\em J. Stat. Phys.} {\bf 64} 541

\bibitem{Step92}
Stepanow S 1992
{\em J. Phys. A: Math. Gen.} {\bf 25} 6187

\bibitem{HV94}
Haronska P and Vilgis T A 1994
{\em J. Chem. Phys.} {\bf 101} 3104

\bibitem{Isich92} 
For a review, see Isichenko M B 1992 
{\em Rev. Mod. Phys.} {\bf 64} 961;
Bouchaud J P and Georges A 1990 {\em Phys. Rep.} {\bf 195} 127

\bibitem{FS97}
Fyodorov Y V and Sommers H-J 1997
{\em J. Math. Phys.} {\bf 38} 1918

\bibitem{SCS88}
Sompolinsky H, Crisanti A and Sommers H-J 1988
{\em Phys. Rev. Lett.} {\bf 61} 259 

\bibitem{HN96}
Hatano N and Nelson D R 1996
{\em Phys. Rev. Lett.} {\bf 77} 570; 1997 {\em Phys. Rev.} B {\bf 56} 8651

\bibitem{IS99}
Izyumov A V and Simons B D 1999 
{\em Europhys. Lett.} {\bf 45} 290

\bibitem{FZ97}
Feinberg J and Zee A 1997 
{\em Nucl. Phys.} B {\bf 501} 643; {\em Nucl. Phys.} B {\bf 504} 579

\bibitem{Edw65}
Edwards S F 1965
{\em Proc. Phys. Soc.} {\bf 85} 613

\bibitem{Klein95}
Kleinert H 1995 
{\em Path Integrals in Quantum Mechanics, Statistics and Polymer Physics}
(Singapore: World Scientific)

\bibitem{SCSS88}
Sommers H-J, Crisanti A, Sompolinsky H and Stein Y 1988
{\em Phys. Rev. Lett.} {\bf 60} 1895

\bibitem{Efetov97}
Efetov K B 1997
{\em Phys. Rev. Lett.} {\bf 79} 491

\bibitem{CW97}
Chalker J T and Wang J 1997
{\em Phys. Rev. Lett.} {\bf 79} 1797

\bibitem{CL95}
See, e.g., Chaikin P M and Lubensky T C 1995
{\em Principles of Condensed Matter Physics}
(Cambridge University Press) 

\bibitem{Lerner99}
Lerner I V, private communication

\bibitem{Lif64}
Lifshitz I M 1964
{\em Usp. Fiz. Nauk} {\bf 83} 617 [1965 {\em Sov. Phys. -- Usp.} {\bf 7} 549];
see also 
Balagurov B Ya and Vaks V G 1974 {\em Sov. Phys. -- JETP} {\bf 38} 968;
Lubensky T C 1984 {\em Phys. Rev.} A {\bf 30} 2657;
Renn S R 1986 {\em Nucl. Phys.} B {\bf 275} 273;
Nieuwenhuizen T M 1989 {\em Phys. Rev. Lett.} {\bf 62} 357
 
\end{references}
\end{document}